\documentclass[twocolumn,english,aps,pra,showpacs,superscriptaddress]{revtex4-1}
\usepackage[T1]{fontenc}
\usepackage[latin9]{inputenc}
\usepackage{amsmath}
\usepackage{amssymb}
\usepackage{stackrel}
\usepackage{graphicx}
\usepackage{esint}

\makeatletter
\usepackage{babel}

\makeatother

\usepackage{babel}
\begin{document}

\title{Quantum Zeno and anti-Zeno effects in open quantum systems}

\author{Zixian Zhou}
\email{zzx1313@sjtu.edu.cn}

\affiliation{Key Laboratory of Artificial Structures and Quantum Control (Ministry
of Education), Department of Physics and Astronomy, Shanghai Jiao
Tong University, Shanghai 200240, China}

\affiliation{Collaborative Innovation Center of Advanced Microstructures, Nanjing
University, Nanjing 210093, China}

\author{Zhiguo L\"{u}}
\email{zglv@sjtu.edu.cn}

\affiliation{Key Laboratory of Artificial Structures and Quantum Control (Ministry
of Education), Department of Physics and Astronomy, Shanghai Jiao
Tong University, Shanghai 200240, China}

\affiliation{Collaborative Innovation Center of Advanced Microstructures, Nanjing
University, Nanjing 210093, China}

\author{Hang Zheng}
\email{hzheng@sjtu.edu.cn}

\affiliation{Key Laboratory of Artificial Structures and Quantum Control (Ministry
of Education), Department of Physics and Astronomy, Shanghai Jiao
Tong University, Shanghai 200240, China}

\affiliation{Collaborative Innovation Center of Advanced Microstructures, Nanjing
University, Nanjing 210093, China}

\author{Hsi-Sheng Goan}
\email{goan@phys.ntu.edu.tw}

\affiliation{Department of Physics and Center for Theoretical Sciences, National
Taiwan University, Taipei 10617}

\affiliation{Center for Quantum Science and Engineering, National Taiwan University,
Taipei 10617, Taiwan}
\begin{abstract}
Traditional approach on quantum Zeno effect (QZE) and quantum anti-Zeno
effect (QAZE) in open quantum systems (implicitly) assumes the bath
(environment) state returning to its original state after each instantaneous
projective measurement on the system and thus ignores the cross-correlations
of the bath operators between different Zeno intervals. However, this
assumption is not generally true, especially for a bath with a considerably
non-negligible memory effect and for a system repeatedly projected
into an initial general superposition state. We find that in stark
contrast to the result of a constant value found in the traditional
approach, the scaled average decay rate in unit Zeno interval of the
survival probability is generally time-dependent or has an oscillatory
behavior. In the case of strong bath correlation, the transition between
the QZE and QAZE depends sensitively on the number of measurements
$N$. For a fixed $N$, a QZE region predicted by the tradition approach
may be in fact already in the QAZE region. We illustrate our findings
using an exactly solvable open qubit system model with a Lorentzian
bath spectral density, which is directly related to realistic circuit
cavity quantum electrodynamics systems. Thus the results and dynamics
presented here can be verified by current superconducting circuit
technology. 
\end{abstract}

\maketitle

\section{Introduction}

With the development of quantum information and computation, quantum
Zeno effect (QZE) \cite{ZenoDef,QZE,expr1,expr2} has attracted much
attention as one of the means to prolong the quantum coherence of
an open quantum system against the influence of its surrounding environment
(bath) \cite{KeepInf1,KeepInf2,KeepEnt}. Another significant effect
in open quantum systems, the quantum anti-Zeno effect (QAZE), i.e.,
if the repeated measurements are not rapid enough, the measurements
may actually enhance the quantum transitions, was revealed by Kofman
and Kurizki \cite{AZE,Kurizki96,expr3}. Each of the repeated measurements
on the system of interest in most studies is considered as an ideal,
instantaneous, projective measurement. Even so, traditional Kofman
and Kurizki approach (KKA) on QZE and QAZE for open quantum systems
(implicitly) assumes the bath state returning to its original state
after each instantaneous projective measurement on the system \cite{ZaZ1,ZaZ2,ZaZ3,Segal2007,Thilagam2010,AZE-MB,Chaudhry2016,Lizuain10,Cao12,Erez08,Kurizki09,Kurizki10E}.
Consequently, the survival probability (SP) $P_{\textrm{KKA}}\left(t\right)$
that the system is still in its initial state $\left|\psi_{S}\right\rangle $
after $N$ repeated measurements with equal time interval $\tau$
is written as 
\begin{eqnarray}
P_{\textrm{KKA}}\left(t\right) & = & \left[P_{\textrm{KKA}}\left(\tau\right)\right]^{N}\nonumber \\
 & = & \left\{ \textrm{Tr}_{S\otimes B}\left[\mathcal{P}_{S}U\left(\tau\right)\rho_{{\rm tot}}\left(0\right)U^{\dagger}\left(\tau\right)\mathcal{P}_{S}
\right]\right\} ^{N},
\end{eqnarray}
where time $t=N\tau$, $P_{\textrm{KKA}}\left(\tau\right)$ is the
SP in the initial state right after a single measurement is performed
($N=1$) \cite{TA_SP_note}, $\mathcal{P}_{S}=\left|\psi_{S}\right\rangle \left\langle \psi_{S}\right|$
is the system state projector, $\rho_{{\rm tot}}\left(0\right)=\mathcal{P}_{S}\otimes\rho_{B}\left(0\right)$
is the initial system-bath state, $U\left(\tau\right)$ is the evolution
operator of the total system-bath Hamiltonian, and $\textrm{Tr}_{S\otimes B}$
denotes taking trace over the degrees of freedoms of the system and
bath. However, this assumption of the KKA is not always valid.

For a general case, the bath state changes throughout the process.
The SP in a general approach (GA) should be 
\begin{equation}
P\left(t\right)=\textrm{Tr}_{S\otimes B}\left\{ \left[\mathcal{P}_{S}U\left(\tau\right)\right]^{N}\rho_{tot}\left(0\right)\left[U^{\dagger}\left(\tau\right)\mathcal{P}_{S}\right]^{N}\right\} ,\label{eq:succ_prob1}
\end{equation}
i.e., the trace over the system and bath variables is performed at
the end of the measurements rather than after each measurement as
in the KKA. In other words, the SP $P_{\textrm{KKA}}\left(t\right)$
in the KKA is just the $N$th power of the SP $P_{\textrm{KKA}}\left(\tau\right)$
and thus neglects the cross-correlation of the bath operators between
different Zeno intervals \cite{ZaZ1,ZaZ2,ZaZ3,Segal2007,Thilagam2010,AZE-MB,Chaudhry2016,Lizuain10,Cao12,Erez08,Kurizki09,Kurizki10E}.
This yields significant quantitative and qualitative different predictions
in QZE and QAZE behaviors between the KKA and the GA, especially when
the repeated measurements project the system into an initial general
superposition state (not just in an initial single excited eigenstate)
and when the bath has a considerably non-negligible memory effect.
It is the aim of this paper to unveil these important differences.
The key qualitative differences we find are as follows. The average
decay rate in each Zeno interval is constant in the KKA, while it
is time-dependent in the GA. In the regime of very small Zeno intervals,
the SP shows exponential-decay behavior in the KKA, but the SP in
the GA can exhibit non-exponential decay. The total average decay
rate depends only on the Zeno interval $\tau$ in the KKA, while it
also depends on the number of repeated measurement $N$ in the GA.
Thus previous studies on QZE-QAZE transitions \cite{ZaZ1,ZaZ2,ZaZ3}
for non-Markovian open quantum systems using the properties of the
total average decay rates need to be reexamined.

\section{Model and dynamics}

We illustrate our results through a qubit system interacting with
a bath that has a non-negligible bath correlation (memory) time \cite{SpinBoson,Breuer02}.
The total Hamiltonian without making the rotating-wave (RW) approximation
in the system-bath coupling reads 
\begin{equation}
H_{\textrm{tot}}=\frac{\Delta}{2}\sigma_{z}+\sum_{k}\omega_{k}b_{k}^{\dagger}b_{k}+g\sigma_{x}\sum_{k}\mu_{k}\left(b_{k}+b_{k}^{\dagger}\right),\label{eq:H_tot}
\end{equation}
where $\sigma_{x,z}$ are the Pauli operators, $b_{k}$ ($b_{k}^{\dagger}$)
is the bath annihilation (creation) operator for bath mode $k$, and
$\Delta$ and $g$ are the qubit frequency and coupling constant,
respectively. We choose the bath spectral density in a Lorentzian
form 
\begin{eqnarray}
J\left(\omega\right) & = & \sum_{k}\left|\mu_{k}\right|^{2}\delta\left(\omega_{k}-\omega\right)\label{eq:muk2}\\
 & = & \frac{\Gamma}{\pi}\frac{1}{\left(\omega-\omega_{0}\right)^{2}+\Gamma^{2}},\label{eq:Jw}
\end{eqnarray}
with width $\Gamma$, central frequency $\omega_{0}$, and normalization
condition $\sum_{k}\mu_{k}^{2}=1$. This not only relates our model
directly to a realistic circuit cavity quantum electrodynamics (QED)
system \cite{Lizuain10,Cao12,Blais04,Wallraff04,Goan,Schoelkopf08,You11,optics},
but also allows a well-defined bath correlation time $1/\Gamma$ to
characterize the memory effect of the bath. Besides, We choose the
initial density matrix for bath as $\rho_{B}\left(0\right)=\left|0_{B}\right\rangle \left\langle 0_{B}\right|$
with bath vacuum $\left|0_{B}\right\rangle $. The Lorentzian bath
initially in the vacuum state $\left|0_{B}\right\rangle $ at zero
temperature makes the spin-boson model with any bilinear form of qubit-bath
coupling (with or without the RW approximation) to be exactly solvable
\cite{hierarchy,MstEq,MstExt}.

\subsection{Bath representation}

We describe next how to obtain an exact evolution equation for the
spin-boson model with a Lorentzian spectral density and any bilinear
form of qubit-bath coupling. First, we discuss how a bath (with many
or infinite degrees of freedom) having a Lorentzian spectral density
can be represented as a single bosonic mode coupling with an interacting
Hamiltonian in a RW form to a fictitious white reservoir \cite{MstExt,diag}.
We show that this representation or decomposition is not an approximation
of the original bath model but rather is exact for bath state initially
in the vacuum state $\left|0_{B}\right\rangle $ at zero temperature.
Consider the qubit-bath (spin-boson) model of Eq. (\ref{eq:H_tot})
in which no RWA is made onto the qubit-bath coupling Hamiltonian.
Suppose we express the bath Hamiltonian consisting of a collection
of an infinite number of harmonic oscillators as 
\begin{equation}
\sum_{k}\omega_{k}b_{k}^{\dagger}b_{k}=\omega_{0}a^{\dagger}a+\sum_{q}\Omega_{q}d_{q}^{\dagger}d_{q}+a^{\dagger}\sum_{q}\gamma_{q}d_{q}+a\sum_{q}\gamma_{q}^{*}d_{q}^{\dagger},\label{HBq}
\end{equation}
where $a$ is the annihilation operator of a single bosonic mode with
characterized frequency $\omega_{0}$, $d_{q}$ is the annihilation
operator of a reservoir mode $q$ with frequency $\Omega_{q}$, and
$\gamma_{q}$ is the coupling strength between the single mode and
the reservoir mode $q$. We may regard the original bath operators
$b_{k}$ as the normal modes of the right-hand-side quadratic RW coupling
model. 

To make this decomposition clearer, let us rewrite the bath Hamiltonian
Eq. (\ref{HBq}) considering the continuous spectrum of excitations
in the bath. Making use of the transformation between the discrete
boson operators $d_{q}$ and the continuous ones $d_{\Omega}$ 
\begin{equation}
d_{q}=\sqrt{D\left(\Omega_{q}\right)}\int_{1/D\left(\Omega_{q}\right)}d\Omega\;d_{\Omega},\label{aktow}
\end{equation}
and a similar transformation between the discrete operators $b_{k}$
and continuous ones $b_{\omega}$, where $D\left(\Omega_{q}\right)d\Omega_{q}$
is the number of modes in the reservoir with frequencies between $\Omega_{q}$
and $\Omega_{q}+d\Omega_{q}$, and $\int_{1/D\left(\Omega_{q}\right)}d\Omega$
represents an integration in a band of width $1/D\left(\Omega_{q}\right)$
around $\Omega_{q}$ \cite{diag}, one obtains 
\begin{eqnarray}
\int\omega b_{\omega}^{\dagger}b_{\omega}d\omega & = & \omega_{0}a^{\dagger}a+\int\Omega d_{\Omega}^{\dagger}d_{\Omega}d\Omega\nonumber \\
 &  & +a^{\dagger}\int\nu_{\Omega}d_{\Omega}d\Omega+a\int\nu_{\Omega}^{*}d_{\Omega}^{\dagger}d\Omega,\label{HBw}
\end{eqnarray}
where $\nu_{\Omega}=\sqrt{D\left(\Omega\right)}\gamma_{\Omega}$,
$\gamma_{\Omega}$ denotes the corresponding quantity of $\gamma_{q}$
in the continuous spectrum representation, and the integral $\int d\Omega=\sum_{q}\int_{1/D\left(\Omega_{q}\right)}d\Omega$
covers the whole spectrum of excitations of the reservoir \cite{diag}.
It has been shown in Ref. \cite{diag} that the Hamiltonian on the
right hand side of Eq. (\ref{HBw}) can be diagonalized and the normal
modes $b_{\omega}$ satisfying $\left[b_{\omega},b_{\omega'}^{\dagger}\right]=\delta\left(\omega-\omega'\right)$
can be expressed as 
\begin{equation}
b_{\omega}=\xi_{\omega}a+\int\eta_{\omega,\Omega}d_{\Omega}d\Omega,\label{bw_NM}
\end{equation}
where $\xi_{\omega}$ and $\eta_{\omega,\Omega}$ satisfy the following
equations 
\begin{equation}
\left|\xi_{\omega}\right|^{2}=\frac{\left|\nu_{\omega}\right|^{2}}{\left[\omega-\omega_{0}-F\left(\omega\right)\right]^{2}+\left[\pi\cdot\left|\nu_{\omega}\right|^{2}\right]^{2}},\label{xiw}
\end{equation}
\begin{equation}
\eta_{\omega,\Omega}=\left[\textrm{P}\frac{1}{\omega-\Omega}+\frac{\omega-\omega_{0}-F\left(\omega\right)}{\left|\nu_{\omega}\right|^{2}}\delta\left(\omega-\Omega\right)\right]\nu_{\Omega}\xi_{\omega},\label{eq:etaw}
\end{equation}
in which $\textrm{P}$ denotes the principle part in the integral,
and 
\begin{equation}
F\left(\omega\right)=\textrm{P}\int\frac{\left|\nu_{\Omega}\right|^{2}}{\omega-\Omega}d\Omega.\label{Fw}
\end{equation}
Furthermore, the single mode $a$ can be re-expressed by the normal
modes as 
\begin{equation}
a=\int f_{\omega}b_{\omega}d\omega.\label{b_NM}
\end{equation}

The coefficient $f_{\omega}$ can be determined as follows. Substituting
Eq. (\ref{b_NM}) for $a$ into the commutator $\left[a,b_{\omega}^{\dagger}\right]$,
one obtains $\left[a,b_{\omega}^{\dagger}\right]=f_{\omega}$; then
substituting Eq.~(\ref{bw_NM}) for $b_{\omega}$ into the same commutator,
one obtains $\left[a,b_{\omega}^{\dagger}\right]=\xi_{\omega}^{*}$.
Thus one concludes the coefficient $f_{\omega}=\xi_{\omega}^{*}$.
The above equations for the diagonalization are all exact and independent
of the expression or form of the spectral density of the reservoir
$d_{\Omega}$. Now, suppose the reservoir is white, i.e., the spectral
density 
\begin{eqnarray}
G\left(\Omega\right) & = & \left|\nu_{\Omega}\right|^{2}\nonumber \\
 & \equiv & \sum_{q}\left|\gamma_{q}\right|^{2}\delta\left(\Omega-\Omega_{q}\right)\nonumber \\
 & = & \Gamma/\pi,\label{nuw}
\end{eqnarray}
then one can easily obtain from Eqs. (\ref{Fw}) and (\ref{xiw})
that $F\left(\omega\right)=0$ and thus 
\begin{equation}
\left|\xi_{\omega}\right|^{2}=\frac{\Gamma}{\pi}\frac{1}{\left(\omega-\omega_{0}\right)^{2}+\Gamma^{2}},
\end{equation}
which is the same Lorentzian form as the spectral density $J\left(\omega\right)$
of Eq. (\ref{eq:Jw}) of the original bath. Consequently, one can,
by making use of Eq. (\ref{b_NM}) with the relation $f_{\omega}=\xi_{\omega}^{*}$
and Eqs. (\ref{eq:muk2}) and (\ref{eq:Jw}), rewrite the single mode
in terms of the normal modes in the discrete form as 
\begin{equation}
a=\sum_{k}\mu_{k}b_{k},\label{a_bk}
\end{equation}
where $b_{k}$ and $\mu_{k}$ are the original bath annihilation operator
and qubit-bath coupling strength, respectively. Furthermore, the commutation
relation $\left[\sum\limits _{k}\mu_{k}b_{k},\sum\limits _{k}\mu_{k}b_{k}^{\dagger}\right]=\sum\limits _{k}\mu_{k}^{2}=1$
confirms once again the relation of Eq. (\ref{a_bk}). Expressing
the original bath modes in the total Hamiltonian (\ref{eq:H_tot})
in terms of the single mode $a$ and the white reservoir modes $d_{q}$,
one obtains
\begin{eqnarray}
H_{\textrm{tot}} & = & \frac{\Delta}{2}\sigma_{z}+g\sigma_{x}\left(a+a^{\dagger}\right)+\omega_{0}a^{\dagger}a+\sum_{q}\Omega_{q}d_{q}^{\dagger}d_{q}\nonumber \\
 &  & +a^{\dagger}\sum_{q}\gamma_{q}d_{q}+a\sum_{q}\gamma_{q}d_{q}^{\dagger},
\end{eqnarray}
where the spectral density of the white reservoir is given by Eq.
(\ref{nuw}). Thus treating the original Lorentzian bath as a single
mode coupled to a flat white reservoir (flat continuum) in a RW form
is an exact result.

\subsection{Exact master equation}

The correlation function of the white-reservoir operators reads 
\begin{eqnarray}
\alpha\left(t,s\right) & = & \sum_{q}\left|\gamma_{q}\right|^{2}e^{-i\Omega_{q}\left(t-s\right)}\nonumber \\
 & = & \int G\left(\Omega\right)e^{-i\Omega\left(t-s\right)}d\Omega\nonumber \\
 & = & \Gamma\delta\left(t-s\right),\label{eq:W_CF}
\end{eqnarray}
that is, the white reservoir correlation time $\tau_{R}\to0$ is treated
as the shortest time scale in the problem. So the degrees of freedom
of the white reservoir can be traced out first regardless of the repeated
projections of the system or the form of the system-bath interaction.

The master equation for the reduced density matrix of a single bosonic
mode (or a harmonic oscillator) coupled to a reservoir (bath) through
a RW-type coupling Hamiltonian can be obtained exactly for an arbitrary
bath spectral density (or bath correlation function) and for an initial
zero-temperature equilibrium reservoirs vacuum state \cite{Str04}
or an initial finite-temperature thermal equilibrium reservoir state
\cite{Yu04}. We consider the original bath state initially in the
zero-temperature vacuum state $\left|0_{B}\right\rangle $, which
translates directly to the no-excitation initial state of $\left|0_{A}\right\rangle \otimes\left|0_{W}\right\rangle $
for the single bosonic mode and the fictitious white reservoir \cite{QPT},
where $\left|0_{A}\right\rangle $ and $\left|0_{W}\right\rangle $
are respectively the vacuum states of the single mode and the fictitious
white reservoir. The exact master equation of Eq. (45) of Ref. \cite{Str04}
was derived using only the condition that the reservoir is initially
in the zero-temperature vacuum state, from which the reservoir's subsequent
evolution to states different from the initial vacuum state can be
determined, and finally the degrees of freedom of the reservoir are
averaged over without any approximation to yield the exact master
equation.

It was also shown in Ref. \cite{Str04} that if the reservoir correlation
function denoted as $\alpha_{CF}\left(t-s\right)$ is replaced by
a $\delta$ function, $\alpha_{CF}\left(t-s\right)=\sum_{\lambda}\left|g_{\lambda}\right|^{2}e^{-i\omega_{\lambda}\left(t-s\right)}=\gamma\delta\left(t-s\right)$,
with some constant $\gamma$, then the exact master equation (36)
or (45) presented in Ref. \cite{Str04} becomes the Lindblad's master
equation in the standard Markov limit. In our case here, the fictitious
white reservoir starts with a reservoir vacuum state $\left|0_{W}\right\rangle $
and has a correlation function delta-correlated in time as in Eq.
(\ref{eq:W_CF}). The constant $\gamma$ used in the correlation function
in Ref. \cite{Str04} equals the twice of the width $\Gamma$ here,
i.e., $\gamma\to2\Gamma$. As a result, we obtain the exact master
equation for the qubit and the single mode here as 
\begin{equation}
\frac{d\tilde{\rho}}{dt}\left(t\right)=\frac{1}{i}\left[H_{\textrm{Rabi}},\tilde{\rho}\left(t\right)\right]-\Gamma\left[a^{\dagger}a\tilde{\rho}\left(t\right)+\tilde{\rho}\left(t\right)a^{\dagger}a-2a\tilde{\rho}\left(t\right)a^{\dagger}\right].\label{eq:exactME}
\end{equation}
Here the qubit-single-mode coupling Hamiltonian 
\begin{equation}
H_{\textrm{Rabi}}=\frac{\Delta}{2}\sigma_{z}+\omega_{0}a^{\dagger}a+g\sigma_{x}\left(a+a^{\dagger}\right),\label{eq:HRabi}
\end{equation}
without the RW approximation is the single-mode version of the spin-boson
Hamiltonian $H_{\textrm{tot}}$ of Eq. (\ref{eq:H_tot}). 

The presence of the Zeno measurements, considered as a series of repeated
projections on the qubit system, does not affect the derived form
of the master equation (\ref{eq:exactME}) when the degrees of freedom
of the fictitious white reservoir is traced out or averaged over \cite{Str04}.
In fact, expressing the projector $\mathcal{P}_{S}=e^{-\mu\left(I-\mathcal{P}_{S}\right)}$
with identity operator $I$ and parameter $\mu\rightarrow+\infty$
\cite{Derivation_Ps}, we can combine the dissipative evolution with
the projective measurement process as a whole non-unitary dynamics
by adding an extra anti-commutator bracket term of $-\mu\cdot C\left(t\right)\ensuremath{\left\{ 1-\mathcal{P}_{S},\rho\right\} }$,
where $C\left(t\right)=\stackrel[n=0]{\infty}{\sum}\delta\left(t-n\tau\right)$
represents a Dirac-comb function. In this representation of Eq. (\ref{eq:exactME}),
the dissipative single (cavity) mode plays the role of the original
bath with a memory time about $1/\Gamma$, and the initial system-bath
state changes from the original bath state $\left|0_{B}\right\rangle $
of $\rho_{\textrm{tot}}\left(0\right)=\mathcal{P}_{S}\otimes\left|0_{B}\right\rangle \left\langle 0_{B}\right|$,
to the single mode state $\left|0_{A}\right\rangle $ of $\tilde{\rho}\left(0\right)=\mathcal{P}_{S}\otimes\left|0_{A}\right\rangle \left\langle 0_{A}\right|$.

We emphasize again that we by no means make the RW approximation on
the qubit-bath coupling Hamiltonian in obtaining Eq. (\ref{eq:exactME})
even though the exact decomposition of the original bath involves
the RW coupling form of a single mode to a white reservoir \cite{MstExt}.
Furthermore, the exact master equation uses only the condition that
the original bath with a Lorentzian bath spectral density is initially
in the zero-temperature vacuum state, or equivalently the fictitious
white reservoir with $\delta$-correlated in time correlation function
is initially in its zero-temperature equilibrium state, i.e., its
vacuum state $\left|0_{W}\right\rangle $. The Lindblad's master equation
(\ref{eq:exactME}) which has the same form as that of a second-order
Markovian master equation is an exact consequence of the model considered
here, rather than a second-order Markovian approximation that assumes
the reservoir correlation time is very short (but not exactly zero,
i.e., correlation function is not really delta-correlated in time)
compared to the other time scales. 

As a result, the evolution within a Zeno interval $\left(n-1\right)\tau<t<n\tau$,
is then determined by Eq.~(\ref{eq:exactME}), and at $t=n\tau$,
the evolution is described by the projective measurement on the system
\begin{eqnarray}
\tilde{\rho}\left(t^{+}\right) & = & \mathcal{P}_{S}\tilde{\rho}\left(t^{-}\right)\mathcal{P}_{S}\nonumber \\
 & = & \mathcal{P}_{S}\otimes\left\langle \psi_{S}\right|\tilde{\rho}\left(t^{-}\right)\left|\psi_{S}\right\rangle .\label{eq:collapse}
\end{eqnarray}
Equation (\ref{eq:collapse}) then serves as the initial state of
Eq.~(\ref{eq:exactME}) for the evolution of the next Zeno interval.
This treatment of the dynamics presented here is exact and only the
initial condition $\rho_{\textrm{tot}}\left(0\right)=\mathcal{P}_{S}\otimes\left|0_{B}\right\rangle \left\langle 0_{B}\right|$
is used to derive the exact master equation even though the Zeno projections
violently change the total state from time to time. That the density
matrix of the original bath will evolve away from the initial vacuum
state $\left|0_{B}\right\rangle \left\langle 0_{B}\right|$ implies
that the density matrix of the single mode will evolve away from its
initial vacuum state $\left|0_{A}\right\rangle \left\langle 0_{A}\right|$.
In addition, the density matrix for the single mode $\left\langle \psi_{S}\right|\tilde{\rho}\left(t^{-}\right)\left|\psi_{S}\right\rangle $
will in general not return to its initial vacuum state $\left|0_{A}\right\rangle \left\langle 0_{A}\right|$
after each measurement. The SP at the final time $t$ is given by
$P\left(t\right)=\textrm{Tr}_{S\otimes A}\tilde{\rho}\left(t\right)$.
In other words, the trace over the bath degrees of freedom (represented
here by the degrees of freedom of the single mode) is performed at
the end of the final time $t$. Our treatment reflects the bath memory
across different Zeno intervals and leads to interesting dynamical
effects. 

\section{Comparison to previous studies}

\begin{figure}
\includegraphics[width=0.9\columnwidth]{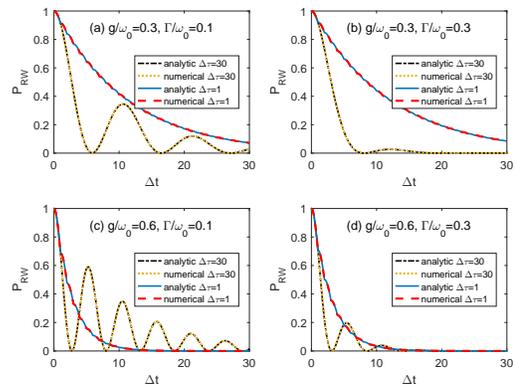}

\caption{\label{fig:RW_e} The SP as functions of time for $\left|\psi_{S}\right\rangle =\left|e\right\rangle $
and $\omega_{0}=\Delta$ with analytical solution from Ref.~\cite{Kurizki96}
and numerical solution from our master equation.}
\end{figure}

\subsection{Coupling Hamiltonian in the RW approximation}

For the population decay model with coupling Hamiltonian in the RW
approximation \cite{Breuer02,RWA,Tai14}, if the measurement in action
is to determine the SP of the excited state $\left|e\right\rangle $
when the initial state is chosen as $\left|e\right\rangle \otimes\left|0_{B}\right\rangle $
\cite{AZE,Kurizki96}. We show next that in this case our master equation
gives the same results of the SP as the exact analytical solutions
given by Ref. \cite{Kurizki96}.

The total Hamiltonian in the RW approximation reads 
\begin{equation}
\ensuremath{H_{\textrm{RW}}=\frac{\Delta}{2}\sigma_{z}+\sum_{k}\omega_{k}b_{k}^{\dagger}b_{k}+g\sum_{k}\mu_{k}\left(\sigma_{+}b_{k}+\sigma_{-}b_{k}^{\dagger}\right)},\label{eq:H_RW}
\end{equation}
where $\sigma_{\pm}$ is the qubit creation/annihilation operator,
respectively. Suppose the bath is initially in the vacuum state $\left|0_{B}\right\rangle $,
then since the total excitation number $N=\sigma_{+}\sigma_{-}+\sum_{k}b_{k}^{\dagger}b_{k}$
of the RW Hamiltonian, Eq.~(\ref{eq:H_RW}), is an invariant quantity
\cite{Breuer02}, the total state at time $t$ for the case of determining
the SP in the excited state $\left|e\right\rangle $ is within the
one-excitation sector and takes the form 
\begin{equation}
\left|\Psi_{{\rm tot}}\left(t\right)\right\rangle =\alpha\left(t\right)\left|e0_{B}\right\rangle +\sum_{k}c_{k}\left(t\right)\left|g1_{k}\right\rangle \label{eq:tot_S}
\end{equation}
with initial condition $\alpha\left(0\right)=1$ and $c_{k}\left(0\right)=0$,
where $\left|1_{k}\right\rangle =b_{k}^{\dagger}\left|0_{B}\right\rangle $
denotes state with one bath boson (photon) in mode $k$. The exact
solution of the time-dependent coefficient $\alpha\left(t\right)$
is given by Eq. (9) of Ref. \cite{Kurizki96} and reads 
\begin{equation}
\alpha\left(t\right)=\frac{1}{2}e^{\left(i\Delta-i\omega_{0}-\Gamma\right)t/2}\left(A_{+}e^{Dt}+A_{-}e^{-Dt}\right)\label{eq:AnaSol}
\end{equation}
with $A_{\pm}=1\pm\left(\Gamma-i\Delta+i\omega_{0}\right)/2D$ and
$D=\sqrt{\frac{1}{4}\left(\Gamma-i\Delta+i\omega_{0}\right)^{2}-g^{2}}$.
So after a Zeno interval $\tau$, the selective measurement to the
qubit excited state projects the total state, Eq. (\ref{eq:tot_S}),
to 
\begin{equation}
\left|\Psi_{{\rm tot}}^{M}\left(\tau^{+}\right)\right\rangle =\left|e\right\rangle \left\langle e|\Psi_{{\rm tot}}\left(\tau^{-}\right)\right\rangle =\alpha\left(\tau\right)\left|e0_{B}\right\rangle ,
\end{equation}
where the superscript $M$ denotes the state it is attached to being
the state right after the measurement, and $\tau^{\pm}$ denote the
times immediately after and before the projective measurement at time
$\tau$, respectively. In other words, the (unnormalized) total state
comes back to its initial form $\left|e0_{B}\right\rangle $ with
additional coefficient $\alpha\left(\tau\right)$, i.e., with survival
probability $P_{\textrm{RW}}\left(\tau\right)=\left|\alpha\left(\tau\right)\right|^{2}$.
The projective measurement removes the system-bath correlation (entanglement)
and the resultant bath state comes back exactly to its initial state
$\left|0_{B}\right\rangle $ after each projective measurement to
the qubit excited state $|e\rangle$. Thus after $n$ Zeno intervals
and $n$ projective measurements to the qubit excited state, one simply
gets the (unnormalized) total state $\left|\Psi_{{\rm tot}}^{M}\left(n\tau^{+}\right)\right\rangle =\alpha\left(n\tau\right)\left|e0_{B}\right\rangle $
with 
\begin{equation}
\alpha\left(n\tau\right)=\left[\alpha\left(\tau\right)\right]^{n}.
\end{equation}
Besides, the survival probability for the qubit to be in the excited
state at $t=n\tau^{+}$ is 
\begin{eqnarray}
P_{\textrm{RW}}\left(n\tau\right) & = & \left|\left\langle e|\Psi_{{\rm tot}}^{M}\left(n\tau^{+}\right)\right\rangle \right|^{2}\nonumber \\
 & = & \left|\alpha\left(n\tau\right)\right|^{2}\nonumber \\
 & = & \left[P_{\textrm{RW}}\left(\tau\right)\right]^{n}.
\end{eqnarray}
The comparison of survival probability $P_{\textrm{RW}}\left(t\right)$
between the above exact analytical solutions \cite{Kurizki96} and
our numerical simulation results using the master equation, Eq. (\ref{eq:exactME}),
with $H_{\textrm{Rabi}}\to H_{\textrm{JC}}=\frac{\Delta}{2}\sigma_{z}+\omega_{0}a^{\dagger}a+g\left(\sigma_{+}a+\sigma_{-}a^{\dagger}\right)$
for the RW coupling Hamiltonian, are presented in Fig. \ref{fig:RW_e}.
One can see that they all coincide with each other for different values
of the coupling constant $g$ and the spectral density width $\Gamma$
(strong coupling case of $g>0.6\omega_{0}$ are also verified although
not shown). In other words, our numerical treatment reproduces exactly
the analytical theory of Ref. \cite{Kurizki96}, regardless of how
large the qubit-bath coupling strength and the bath correlation time
are. This fact demonstrates that our master equation is exact (even
though the white noise dissipative terms look like an standard second-order
Markovian Lindblad equation), and thus our master equation approach
is a correct and valid tool to study the qubit-bath dynamics in the
quantum Zeno process.

\begin{figure}
\includegraphics[width=0.9\columnwidth]{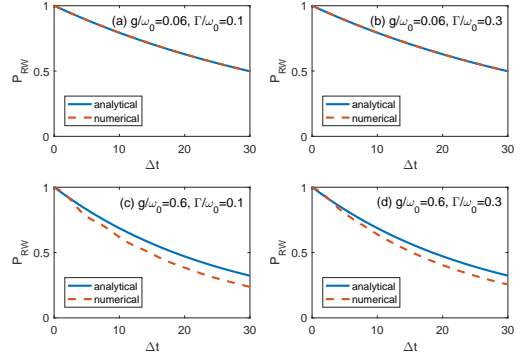}

\caption{\label{fig:RW_gen}The SP as functions of time with the approximated
analytical result and our master equation, in which Zeno interval
$\Delta\tau=0.1$, $\omega_{0}=\Delta$ and $\left|\psi_{S}\right\rangle =0.8\left|e\right\rangle +0.6\left|g\right\rangle $. }
\end{figure}

Actually only in the above case of determining the SP in the excited
state $\left|e\right\rangle $ is the result of SP the GA the same
as that in the KKA \cite{Kurizki96}. However, if the repeated measurements
project the qubit system into an initial general superposition state
of $\left|\psi_{S}\right\rangle =\alpha\left|e\right\rangle +\beta\left|g\right\rangle $
(not just into the initial excited state $\left|\psi_{S}\right\rangle =\left|e\right\rangle $),
where $\left|g\right\rangle $ is the qubit ground state, the bath
state after each projective measurement is different. Within the first
Zeno interval $0<t<\tau$, the total state can be written as 
\begin{equation}
\left|\Psi_{{\rm tot}}\left(t\right)\right\rangle =\alpha\left(t\right)\left|e0_{B}\right\rangle +\beta\left(t\right)\left|g0_{B}\right\rangle +\sum_{k}c_{k}\left(t\right)\left|g1_{k}\right\rangle 
\end{equation}
with the initial condition $\alpha\left(0\right)=\alpha$, $\beta\left(0\right)=\beta$
and $c_{k}\left(0\right)=0$. The time-dependent coefficients can
still be exactly obtained within the first Zeno interval with $\alpha\left(t\right)$
given by Eq. (\ref{eq:AnaSol}) and $\beta\left(t\right)=\beta$.
Then the projection of the selective measurement at time $\tau$ makes
the (unnormalized) bath state be 
\begin{eqnarray}
\left|\psi_{B}^{M}\left(\tau^{+}\right)\right\rangle  & = & \left\langle \psi_{S}|\Psi_{{\rm tot}}\left(\tau\right)\right\rangle \nonumber \\
 & = & \left[\alpha^{\ast}\alpha\left(\tau\right)+\left|\beta\right|^{2}\right]\left|0_{B}\right\rangle +\beta^{\ast}\sum_{k}c_{k}\left(\tau\right)\left|1_{k}\right\rangle .
\end{eqnarray}
One clearly sees that this bath state does not return to the initial
bath vacuum state $\left|0_{B}\right\rangle $. The SP after the first
measurement can be calculated exactly as 
\begin{eqnarray}
P_{\textrm{RW}}\left(\tau\right) & = & \left\langle \psi_{B}^{M}\left(\tau^{+}\right)|\psi_{B}^{M}\left(\tau^{+}\right)\right\rangle \nonumber \\
 & = & \left|\alpha^{\ast}\alpha\left(\tau\right)+\left|\beta\right|^{2}\right|^{2}+\left|\beta\right|^{2}\sum_{k}\left|c_{k}\left(\tau\right)\right|^{2}\nonumber \\
 & = & \left|\alpha^{\ast}\alpha\left(\tau\right)+\left|\beta\right|^{2}\right|^{2}+\left|\beta\right|^{2}\left(\left|\alpha\right|^{2}-\left|\alpha\left(\tau\right)\right|^{2}\right)\label{eq:P_tau}
\end{eqnarray}
However, the (unnormalized) initial total qubit-bath state for the
second Zeno interval reads 
\begin{eqnarray}
\left|\Psi_{{\rm tot}}^{M}\left(\tau^{+}\right)\right\rangle  & = & \left|\psi_{S}\right\rangle \otimes\left|\psi_{B}^{M}\left(\tau^{+}\right)\right\rangle \nonumber \\
 & = & \left|\psi_{S}\right\rangle \otimes\left[\alpha^{\ast}\alpha\left(\tau\right)+\left|\beta\right|^{2}\right]\left|0_{B}\right\rangle \nonumber \\
 &  & +\left|\beta\right|^{2}\sum_{k}c_{k}\left(\tau\right)\left|g\right\rangle \otimes\left|1_{k}\right\rangle \nonumber \\
 &  & +\beta^{\ast}\alpha\sum_{k}c_{k}\left(\tau\right)\left|e\right\rangle \otimes\left|1_{k}\right\rangle ,\label{Psi_M_G}
\end{eqnarray}
which contains a two-excitation state $\left|e1_{k}\right\rangle $
that goes out the zero-excitation and one-excitation Hilbert space
that we set initially for the total state evolution in the first Zeno
interval. Continuing the analysis, one finds the initial total state
for the evolution of the $n$-th Zeno interval contains $n$ excitations,
which is too complex to solve analytically. Thus, we have $P_{\textrm{RW}}\left(n\tau\right)\neq\left[P_{\textrm{RW}}\left(\tau\right)\right]^{n}$
for a general initial qubit state even with the qubit-bath coupling
Hamiltonian in the RW approximation. In the following, we still compare
the SP in this case between the approximated analytical result $P_{\textrm{RW}}\left(n\tau\right)\approx\left[P_{\textrm{RW}}\left(\tau\right)\right]^{n}$
which assumes the bath state return back to its initial state after
each projective measurement with that of our master equation in Fig.
\ref{fig:RW_gen}. For weak-coupling case of $g=0.06$ shown in Figs.
\ref{fig:RW_gen} (a) and \ref{fig:RW_gen} (b), the approximated
results of the KKA \cite{Kurizki96} agree very well with our exact
numerical results, while they deviate from each other in the strong
coupling case of $g=0.6$ as shown in Figs. \ref{fig:RW_gen} (c)
and \ref{fig:RW_gen} (d). The deviation certainly comes from the
changes of the bath state. Therefore, our calculated results demonstrate
that the bath state indeed changes in the Zeno projection process
in the strong coupling regime.

\subsection{Coupling Hamiltonian without the RW approximation}

\begin{figure}
\noindent \includegraphics[width=0.9\columnwidth]{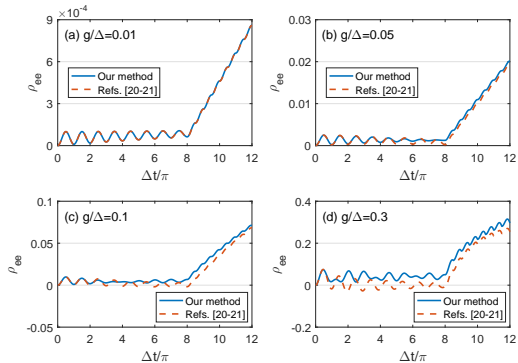}

\caption{\label{fig:non-selct} Excited-level populations given by our master
equation (blue solid lines) and the master (rate) equation used in
Refs. \cite{Erez08,Kurizki09} (red dashed lines). The dynamic from
$\Delta t=0$ to $\Delta t=8\pi$ corresponds to natural relaxation,
and for $\Delta t>8\pi$, it experiences non-selective measurements
with interval $\Delta\tau=\pi/2$. Parameters are $\omega_{0}=\Delta$,
$\Gamma/\Delta=0.03$. }
\end{figure}

For the original spin-boson model without the RW approximation, one
cannot obtain an exact solution in the total wave function approach
even for the first Zeno interval. References \cite{Erez08,Kurizki09}
studied the spin or qubit under repeated non-selective quantum non-demolition
(QND) measurements in this model using a perturbation theory in system-bath
coupling strength and assumed the bath as an immutable entity. The
effect of non-selective non-intrusive QND measurements is to erase
the qubit-bath correlation, transforming their joint density matrix
into an approximated factorized form. Then the reduced density matrix
of the qubit remains diagonal throughout the considered evolution
and can always be written in Gibbs form $\rho_{S}\left(t\right)=Z^{-1}e^{-\beta\left(t\right)H_{S}}$,
where $\beta\left(t\right)$ is the time-dependent effective inverse
temperature that characterizes ``heating'' and ``cooling'' (quoted
from Sec. 2.1 and 2.2 of Ref. \cite{Kurizki09}).

Despite the existing key differences in measurement scenario and in
measurement effect on the subsequent qubit dynamics, we make comparisons
and clarify the validity between the master (rate) equation used in
Refs. \cite{Erez08,Kurizki09} and that in our work. Taking the zero-temperature
system-bath product state $\left|g0_{B}\right\rangle $ as the initial
state (which is the same as that in Figure 1 of Ref. \cite{Erez08}),
where $\left|0_{B}\right\rangle $ represents the bath vacuum state,
we calculate the excited-state population $\rho_{ee}$ by our master
equation following the dynamical rules in Ref. \cite{Erez08}. References
\cite{Erez08,Kurizki09} provided equations of motion of the elements
of the reduced density matrix 
\begin{equation}
\frac{d}{dt}\rho_{ee}=-\frac{d}{dt}\rho_{gg}=-R_{e}\rho_{ee}+R_{g}\rho_{gg}\label{eq:rateE}
\end{equation}
with $R_{e}\left(t\right)=2\int G_{0}\left(\omega\right)\frac{\sin\left(\omega-\Delta\right)t}{\omega-\Delta}d\omega$
and $R_{g}\left(t\right)=2\int G_{0}\left(\omega\right)\frac{\sin\left(\omega+\Delta\right)t}{\omega+\Delta}d\omega$.
Taking the Lorentzian spectrum to be 
\begin{equation}
G_{0}\left(\omega\right)=g^{2}\frac{1}{\pi}\frac{\Gamma}{\left(\omega-\omega_{0}\right)^{2}+\Gamma^{2}},
\end{equation}
we present in Fig. \ref{fig:non-selct} the excited-state population
as a function of time given by the two different master (rate) equations,
namely, Eqs.(\ref{eq:exactME}) and (\ref{eq:rateE}). One can see
in the weak-coupling regime (Figs. \ref{fig:non-selct} (a) and \ref{fig:non-selct}
(b)), the results obtained by the two different master (rate) equations
agree well with each other. It demonstrates that our master equation
can reproduce the same heating-up behaviors studied in Ref. \cite{Erez08}.
However, in the cases of moderate coupling (Figs. \ref{fig:non-selct}
(c) and \ref{fig:non-selct} (d)), the results by the two master equations
are significantly different in the large-time regime, in which the
excited-state population in red-dashed lines given by the master (rate)
equation of Refs. \cite{Erez08,Kurizki09} even fall below zero. This
nonphysical result which is more evident in the strong-coupling regime
indicates that the master (rate) equation in Refs. \cite{Erez08,Kurizki09}
becomes improper to use in these cases. While our exact master equation
is still suitable even in the strong-coupling regime.

We note here that in the main text and Supplementary Information in
Ref. \cite{Erez08}, the post-measurement bath state and the system-bath
correlations are described both analytically and numerically, and
in the Supplementary Information of Ref. \cite{,Kurizki09} the small
deviation of the bath state from the original Gibbs form was discussed
in the weak-coupling perturbation theory, whereas Ref. \cite{Kurizki10E}
shows that the bath change is drastic if only few modes in the bath
play a role. Theses studies \cite{Erez08,Kurizki09,Kurizki10E} recognized
changes of bath state, but the effects were argued not to be substantial
due mainly to the fact that many or an infinite number of bath modes
were considered and the investigations were conducted within the weak-coupling
perturbation theory \cite{Erez08,Kurizki09}. By using our approach
of representing the infinite number of modes of the original Lorentzian
bath as a single mode coupled to a fictitious white reservoir of an
infinite number of modes, then after the infinite number of modes
of the white reservoir are traced out, the resultant master equation
describe a qubit interacting with effectively a dissipative single
mode. When only one bath mode plays a significant role, the results
of Refs. \cite{Erez08,Kurizki09,Kurizki10E} will also apply to this
case of bath changes.

\section{Effects of bath state changes and bath correlation time on QZE}

Next we analyze the properties of the average decay rate in each Zeno
interval defined by $\lambda_{n}=\frac{1}{\tau}\ln\left[P\left(n\tau\right)/P\left(n\tau+\tau\right)\right]$.
As stated, the bath state after a projective measurement for general
situations and models is different from the bath state after its previous
measurement (i.e., the initial state at the beginning of each Zeno-interval
evolution is different), and thus the average decay rates in different
Zeno intervals do not equal to each other, which display rich effects
and phenomena. To characterize the changing decay rates between different
Zeno intervals, we investigate the behavior of the average decay rate
in each interval $\lambda_{n}$. In the KKA (or in the RW-approximated
model with projection measurement into $\left|\psi_{S}\right\rangle =\left|e\right\rangle $
in Ref. \cite{AZE,Kurizki96}), only the total average decay rate
$\Lambda_{N}\left(\tau\right)=-\ln P\left(N\tau\right)/N\tau$ is
used due to the assumption (fact) that the bath state does not change
from its initial state and the average decay rates in different Zeno
intervals are the same (i.e., the total average decay rate equals
to the average decay rate in a single Zeno interval). Furthermore,
the QZE ($\lambda_{n}\to0$ as $\tau\to0$) indicates $\lambda_{n}\propto\tau$
for small $\tau$, so it is natural to define their ratio $w_{n}=\lambda_{n}/\tau$
as a meaningful and significant physical quantity to characterize
the general QZE. We call $w_{n}$ the scaled decay rate in unit Zeno
interval. In the limiting case of continuous Zeno measurements in
which $\tau\rightarrow0$, the ratio $w_{n}$ is actually finite and
the discrete series $w_{n}$ becomes a continuous function of time,
namely, 
\[
\lim_{\tau\to0,n\tau\to t}\left(\lambda_{n}/\tau\right)=w\left(t\right),
\]
and the SP takes the form of 
\begin{equation}
P\left(t\right)=\exp\left[-\tau\int_{0}^{t}w\left(t'\right)dt'\right].\label{eq:surv}
\end{equation}

When $w\left(t\right)$ is a constant, the decay is exponential. But
if $w\left(t\right)$ varies explicitly with time, the decay is non-exponential.
$w\left(t\right)$ in the KKA is always a constant. Thus in the Zeno
limit of $\tau\to0$, the SP always shows exponential-decay behavior
in the KKA, but the SP in the GA can still exhibit non-exponential
decay. We can derive an analytical expression for the SP in the $\tau\to0$
limit, which not only can provide us with an understanding of SP in
the very short $\tau$ regime but also give a verification of the
numerical master equation approach. To obtain the explicit analytical
expression of $w\left(t\right)$ in the $\tau\to0$ limit, we first
directly calculate the total state $\left|\Psi_{\textrm{tot}}\left(t\right)\right\rangle $
after $n$ measurements by $\left|\Psi_{\textrm{tot}}\left(t\right)\right\rangle =\left(\mathcal{P}_{S}e^{-iH_{\textrm{tot}}\tau}\right)^{n}\left|\psi_{S}0_{B}\right\rangle $.
Then obtaining the SP $P\left(t\right)=\left\langle \Psi_{\textrm{tot}}\left(t\right)\right|\mathcal{P}_{S}\left|\Psi_{\textrm{tot}}\left(t\right)\right\rangle $
to the dominant order in $\tau$ and expressing it in the form of
Eq. (\ref{eq:surv}), we obtain 
\begin{equation}
w\left(t\right)=\left\langle H_{S\eta}^{2}\left(t\right)\right\rangle -\left\langle H_{S\eta}\left(t\right)\right\rangle ^{2}+g^{2}\left(1-\left\langle \sigma_{x}\right\rangle {}^{2}\right),\label{eq:w_xpr}
\end{equation}
where $H_{S\eta}\left(t\right)=\Delta\sigma_{z}/2+g\sigma_{x}\left[\eta\left(t\right)+\eta^{\ast}\left(t\right)\right]$,
and real function 
\begin{equation}
\eta\left(t\right)=g\left\langle \sigma_{x}\right\rangle \left[e^{-\left(\Gamma+i\omega_{0}\right)t}-1\right]/\left(\omega_{0}-i\Gamma\right).\label{eq:eta}
\end{equation}
This analytical expression of Eqs. (\ref{eq:surv}) and (\ref{eq:w_xpr}),
provides a good check for the SP in the small $\tau$ regime calculated
by the numerical method of Eqs. (\ref{eq:exactME}) and (\ref{eq:collapse}).

\begin{figure}
\includegraphics[width=0.9\linewidth]{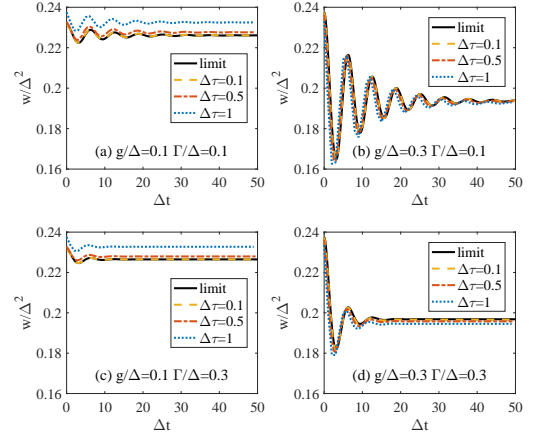} \caption{\label{fig:w_fit}(color online). Scaled decay rate per Zeno interval
$w_{n}=\lambda_{n}/\tau$ as functions of time $t$ with Zeno interval
$\tau=1/\Delta$ (blue dotted lines), $\tau=0.5/\Delta$ (red dash-dot
lines), and $\tau=0.1/\Delta$ (yellow dashed lines) for different
values of $\Gamma$ and $g$ by the numerical solutions. The black
solid lines are the analytical results in the continuous limit $\tau\rightarrow0$
from Eq. (\ref{eq:w_xpr}). In subgraph (a) and (b), $\Gamma/\Delta=0.1$;
in subgraph (c) and (d), $\Gamma/\Delta=0.3$. In subgraph (a) and
(c), $g/\Delta=0.1$; in subgraph (b) and (d), $g/\Delta=0.8$. The
initial states is $\left|\psi_{S}\right\rangle =3/5\left|e\right\rangle +4/5\left|g\right\rangle $
with $\left|e\right\rangle $ and $\left|g\right\rangle $ being the
ground and excited states of the qubit, and the parameter $\omega_{0}/\Delta=1$.}
\end{figure}

In Fig. \ref{fig:w_fit}, the numerical results of $w_{n}$ along
with the analytical result $w\left(t\right)$ are presented for repeated
projections to a general initial system state but different $\tau$,
$g$ and $\Gamma$. The series $\left\{ \lambda_{n}=w_{n}\tau\right\} $
that has an oscillatory behavior as a function of $n$ ($t=n\tau$
for a fixed $\tau$) refers to the variation of the average decay
rate cross different Zeno intervals, which is significantly different
from the oscillatory behavior of the SP (not in the average decay
rate) over time $t$ obtained by the KKA or presented in Refs. \cite{AZE,Kurizki96}
in which the average decay rate in each Zeno interval is a constant.
The numerical results for short $\tau=0.1/\Delta$ (yellow dashed
lines) agree quite well with the analytical ones (black solid lines).
Besides, $w\left(t\right)$ exhibits a damped oscillation with time,
indicating that the scaled average decay rate in unit $\tau$ for
a general initial state is qualitatively different from the constant
average decay rate of the traditional QZE.

Furthermore, Fig. \ref{fig:w_fit} presents the quantitative effects
of the bath central frequency $\omega_{0}$, the qubit-bath coupling
strength $g$, and the bath memory time $1/\Gamma$ on the non-exponential
decay of $P\left(t\right)$ through the behavior of $w\left(t\right)$.
In each subgraph, $w\left(t\right)$ clearly exhibits damped oscillations
because $w\left(t\right)$ of Eq. (\ref{eq:w_xpr}) contains both
$\eta\left(t\right)+\eta^{\ast}\left(t\right)$ and its square term
with damped oscillation frequencies $\omega_{0}$ and $2\omega_{0}$,
respectively. Since the $2\omega_{0}$ term is proportional to $g^{4}$,
its contribution is much less than the $\omega_{0}$ term that is
proportional to $g^{2}$ for small coupling strengths. Thus the $2\omega_{0}$
component visible in Fig. \ref{fig:w_fit} (b) is not seen in Fig.
\ref{fig:w_fit} (a). Moreover, as $g$ decreases from Fig. \ref{fig:w_fit}
(b) to (a) as well as Fig. \ref{fig:w_fit} (d) to (c), the amplitudes
of the damped oscillations also decrease. This indicates that for
very small system-bath coupling the KKA to assume the bath state does
not changes significantly from its original state can be justified
\cite{Erez08,Kurizki09,Kurizki10E}. Figure \ref{fig:w_fit} also
shows the influence of $\Gamma$ on the damping behavior of $w\left(t\right)$.
The damping rate of $w\left(t\right)$ is, as shown in Eq. (\ref{eq:eta}),
just the width $\Gamma$ of the Lorentzian spectrum, namely, the dissipation
rate of the single (cavity) mode, whose inverse value $1/\Gamma$
characterizes the memory time of the Lorentzian bath. When $\Gamma=0$,
the qubit is effectively coupled to a single mode and exchange information
with it periodically. As a result, $w\left(t\right)$ oscillates without
damping. For finite values of $\Gamma$, if the evolution time $t$
is much larger than the memory time $1/\Gamma$, then $w\left(t\right)$
will approach a constant value just like the traditional QZE.

\begin{figure}
\includegraphics[width=0.9\linewidth]{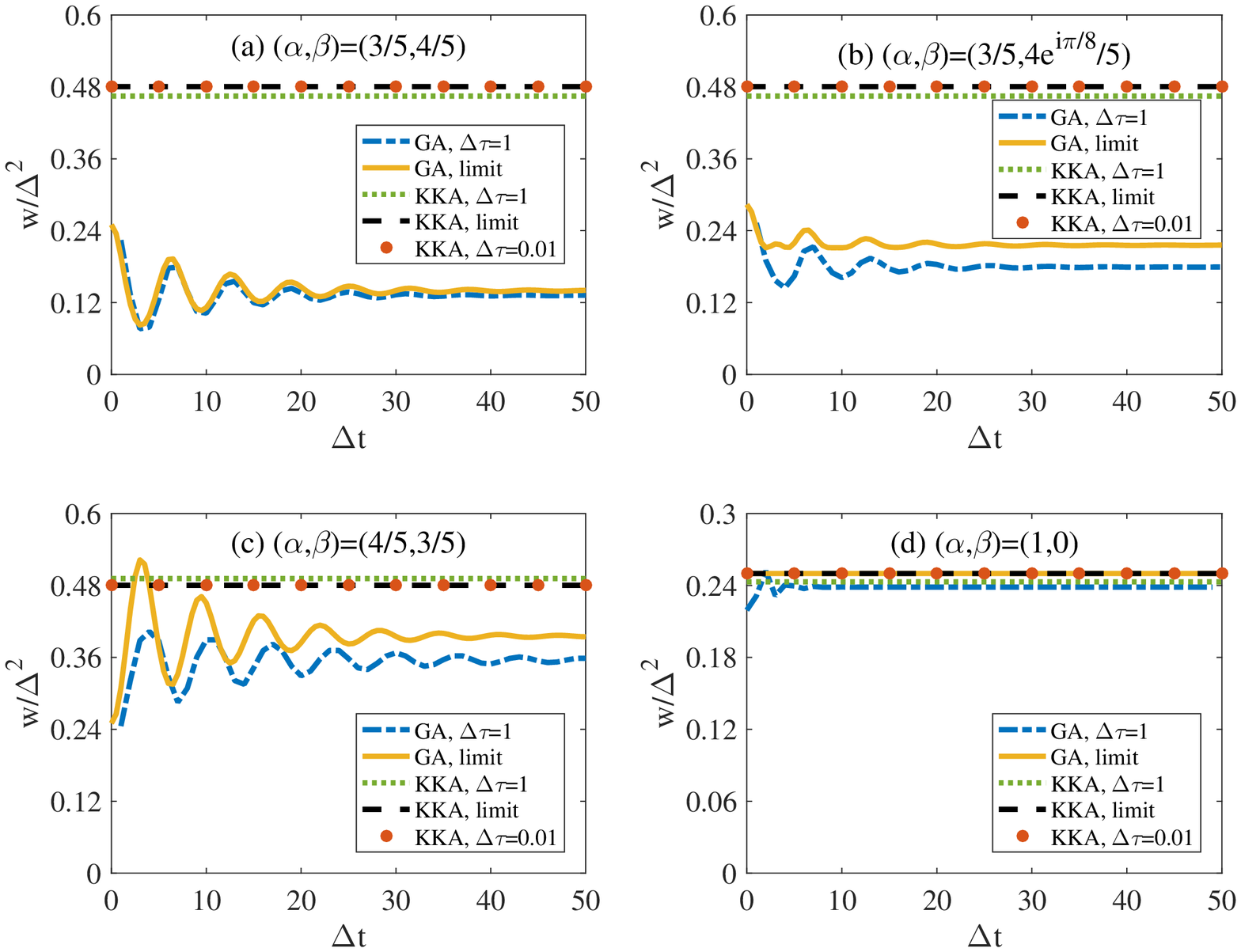} \caption{\label{fig:w_compare}(color online). Functions $w\left(t\right)$
and constants $w_{\textrm{KKA}}$ for different initial states of
$\left|\psi_{S}\right\rangle =\alpha\left|e\right\rangle +\beta\left|g\right\rangle $
with $\left(\alpha,\beta\right)$ equal to (a) $\left(3/5,4/5\right)$,
(b) $\left(3/5,e^{i\pi/8}4/5\right)$, (c) $\left(4/5,3/5\right)$,
and (d) $\left(1,0\right)$. The results of $w\left(t\right)$ for
$\tau=1/\Delta$ (blue dash-dotted lines) are calculated numerically
using Eqs. (\ref{eq:exactME}) and (\ref{eq:collapse}) and for the
continuous limit $\tau\rightarrow0$ (yellow solid lines) calculated
analytically using Eq. (\ref{eq:w_xpr}). The results of $w_{\textrm{KKA}}$
for $\tau=1/\Delta$ (green dotted lines) as well as $\tau=0.01/\Delta$
(red dots) are calculated numerically and for the continuous limit
$\tau\rightarrow0$ (black dashed lines) calculated analytically using
the formulas in the KKA described in the main text. Other parameters
used are $\omega_{0}/\Delta=1$, $g/\Delta=0.5$, and $\Gamma/\Delta=0.1$.}
\end{figure}

The analytical result of $w_{\textrm{KKA}}$ in the continuous limit
of $\tau\to0$ in the KKA can be found by $w_{\textrm{KKA}}=\left\langle \Psi_{\textrm{tot}}\left(t\right)\right|H_{\textrm{tot}}^{2}\left|\Psi_{\textrm{tot}}\left(t\right)\right\rangle -\left\langle \Psi_{\textrm{tot}}\left(t\right)\right|H_{\textrm{tot}}\left|\Psi_{\textrm{tot}}\left(t\right)\right\rangle ^{2}=\left(\Delta/2\right)^{2}\left(1-\left\langle \sigma_{z}\right\rangle ^{2}\right)+g^{2}$
\cite{AZE}. For finite Zeno interval $\tau$, we can express the
SP $P_{\textrm{KKA}}\left(\tau\right)$ associated with one measurement
in the KKA as $P_{\textrm{KKA}}\left(\tau\right)=\left|\left\langle \psi_{S}0_{A}\right|e^{-iH_{\textrm{eff}}\tau}\left|\psi_{S}0_{A}\right\rangle \right|^{2}$,
where $H_{\textrm{eff}}=H_{\textrm{Rabi}}-i\Gamma a^{\dagger}a$ is
the effective non-Hermitian Hamiltonian that takes into account the
single mode decay \cite{MstExt,Gardiner00}. The result of $w_{\textrm{KKA}}$
for finite $\tau$ can thus be obtained by $w_{\textrm{KKA}}=-\frac{1}{\tau^{2}}\ln P_{\textrm{KKA}}\left(\tau\right)$
with the dynamics of $P_{\textrm{KKA}}\left(\tau\right)$ solved numerically.
The comparisons between functions $w\left(t\right)$ and constants
$w_{\textrm{KKA}}$ for the same parameters but different values of
$\tau$ are presented in each subgraph of Fig. \ref{fig:w_compare}.
One can see that the numerical results of $w_{\textrm{KKA}}$ for
$\Delta\tau=0.01$ (red dots) agree well with the analytical results
of $w_{\textrm{KKA}}$ for $\tau\to0$ (black dashed lines), which
verifies again the single-mode approach used in this paper. Compared
to $w_{\textrm{KKA}}$, the function $w\left(t\right)$ taking account
of the cross-correlation of the bath operators between different Zeno
intervals and the bath memory time exhibits rich phenomena. The SP
right after the first Zeno measurement of the GA is always larger
than or equal to that of the KKA since $P\left(\tau\right)=\textrm{Tr}_{B}\left\langle \psi_{S}\right|\rho_{\textrm{tot}}\left(\tau\right)\left|\psi_{S}\right\rangle \geq\left\langle 0_{B}\right|\left\langle \psi_{S}\right|\rho_{\textrm{tot}}\left(\tau\right)\left|\psi_{S}\right\rangle \left|0_{B}\right\rangle =P_{\textrm{KKA}}\left(\tau\right)$.
Note again that in the Zeno limit of $\tau\to0$, $w_{\textrm{KKA}}$
is a constant but $w\left(t\right)$ shows damped oscillation behavior
for general initial states. The constant $w_{\textrm{KKA}}$, by means
of Eq. (\ref{eq:surv}), leads to the exponential-decay SP $P\left(t\right)=e^{-\tau w_{\textrm{KKA}}t}$
($w_{\textrm{KKA}}^{-1/2}$ is just the Zeno time). Therefore, the
fact that the SP of a general initial qubit state in the regime of
very small Zeno intervals shows exponential-decay behavior in the
KKA but shows non-exponential decay in our GA, is also an important
major difference between these two different approaches, even though
at large Zeno time intervals the different approaches may all show
damped oscillatory behaviors in the SP. Depending on the initial states
and the value of $\tau$, $w\left(t\right)$ can then, as shown in
Fig. \ref{fig:w_compare}, be larger or less than $w_{\textrm{KKA}}$.

Moreover, the relative phase between the basis states of $\left|e\right\rangle $
and $\left|g\right\rangle $ of the initial qubit state $\left|\psi_{S}\right\rangle =\alpha\left|e\right\rangle +\beta\left|g\right\rangle $
has, by comparing Fig. \ref{fig:w_compare} (a) with Fig. \ref{fig:w_compare}
(b), an important effect on $w\left(t\right)$. In contrast, the TR
results near the continuous limit do not depend on the relative phase
in the initial state, for the initial phase is not explicitly contained
in the expression of $w_{\textrm{KKA}}$. In fact, in the model investigated,
$w_{\textrm{KKA}}$ in the continuous limit depends only on $1-\left\langle \sigma_{z}\right\rangle ^{2}$,
so $w_{\textrm{KKA}}$ is the same for the particularly chosen different
initial states in Figs. \ref{fig:w_compare} (a), (b), and (c). In
Fig. \ref{fig:w_compare} (d), the scaled decay rates $w\left(t\right)$
in the continuous limit of $\tau\rightarrow0$ (yellow solid line)
is a constant and equals to $w_{\textrm{KKA}}$ (black dashed line),
i.e., $w\left(t\right)=w_{\textrm{KKA}}=g^{2}$. This is because the
initial state $\left|\psi_{S}\right\rangle =\left|e\right\rangle $
makes $\left\langle \sigma_{x}\right\rangle =0$ and thus according
to Eq. (\ref{eq:eta}) the amplitudes of the oscillation parts of
$w\left(t\right)$ are zero. But for finite Zeno interval $\tau$,
$w_{n}$ still oscillates with time (blue dash-dot line), attributed
to the higher-order effect in finite value of $\tau$.

\section{Transition between QZE and QAZE}

Next we discuss the transition between QZE and QAZE. It is known that
longer Zeno interval may lead to the QAZE. In the KKA, the total average
decay rate $\lambda_{\textrm{KKA}}\left(\tau\right)=-\ln P\left(t\right)/t=-\ln P\left(\tau\right)/\tau$
depends only on $\tau$ and is independent of the number of measurements
$N$. One may define $\frac{d}{d\tau}\lambda_{\textrm{KKA}}\left(\tau\right)>0$
as the QZE, and $\frac{d}{d\tau}\lambda_{\textrm{KKA}}\left(\tau\right)<0$
as the QAZE, with the QZE-QAZE transition point called the transition
time $\tau^{c}$ \cite{AZE,Segal2007,Thilagam2010,AZE-MB,Chaudhry2016}.
As we have seen, the decay rate per Zeno interval $\lambda_{n}$ or
scaled decay rate $w_{n}$ varies also with the number of measurements.
Thus the QZE-AZE transition point should depend also on the number
of measurements $N$ \cite{AZE-MB}. The total average decay rate,
$\Lambda_{N}\left(\tau\right)$, for $N$ measurements is defined
as $\Lambda_{N}\left(\tau\right)=-\ln P\left(N\tau\right)/N\tau=\frac{1}{N}\sum_{n=0}^{N-1}\lambda_{n}$.
For each given $N$, we define $\frac{d}{d\tau}\Lambda_{N}\left(\tau\right)>0$
as the QZE, and $\frac{d}{d\tau}\Lambda_{N}\left(\tau\right)<0$ as
the QAZE, with the transition time $\tau_{N}^{c}$ given by the transition
points. This definition for the transition between QZE and QAZE is
a straightforward extension of the traditional definition, i.e., for
$N=1$, it goes back to the traditional definition.

\begin{figure}
\includegraphics[width=0.9\linewidth]{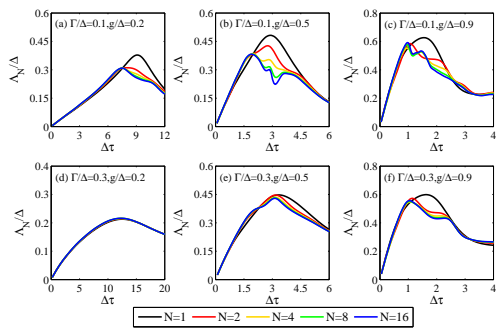}\caption{\label{fig:QAZE}(color online). Total average decay rate $\Lambda_{N}$
as a function of Zeno interval $\tau$ for various number of measurements
$N$. The width in (a), (b) and (c) is $\Gamma/\Delta=0.1$ and in
(d), (e) and (f) is $\Gamma/\Delta=0.3$. The coupling strength in
(a) and (d) is $g/\Delta=0.2$, in (b) and (e) is $g/\Delta=0.5$,
and in (c) and (f) is $g/\Delta=0.9$. The black, red, yellow, green,
and blue lines correspond to $N=1,\ 2,\ 4,\ 8,\ 16$, respectively.
The initial state is $\left|\psi_{S}\right\rangle =\left|e\right\rangle $
and the other parameter used is $\omega_{0}/\Delta=1$.}
\end{figure}

The total average decay rates $\Lambda_{N}\left(\tau\right)$ as functions
of $\tau$ with initial state $\left|\psi_{S}\right\rangle =\left|e\right\rangle $
for various $N$ presented in Fig. \ref{fig:QAZE} are different from
each other, and for each $N$ there is a corresponding transition
time $\tau_{N}^{c}$. This is qualitatively different from the traditional
QZE-QAZE transition. As $N$ increases, the transition time $\tau_{N}^{c}$
becomes smaller. This may lead to an interesting result. For example,
in Fig. \ref{fig:QAZE} (b), the transition point for $N=1$ (black
line) is near $\tau=3/\Delta$, while those for $N=8,16$ (green and
blue lines) are close to $\tau=2/\Delta$. If the Zeno interval is
set to be fixed at $\tau=2.5/\Delta$, then the KKA ($N=1$) would
predict a QZE while the general approach predicts a qualitatively
different QAZE for large $N$ measurements. Besides, the blue curves
in Fig. \ref{fig:QAZE} (b) and (c) show multi transition points,
which might be regarded as multi QZE-QAZE transitions \cite{AZE-MB}.

The parameters $\Gamma$ and $g$ also have significant effects on
the transition between QZE and QAZE. In Fig. \ref{fig:QAZE} (a),
(b), and (c) with a small width $\Gamma/\Delta=0.1$, the curves for
various $N$ separate from one another, while in Figs. \ref{fig:QAZE}
(d), (e), and (f) with a larger width $\Gamma/\Delta=0.3$, the curves
almost overlap with one another. One can also observe that the curves
of various $N$ deviate from one another in the strong coupling regime
(Fig. \ref{fig:QAZE} (c) and (f)) but are close to one another in
the weak coupling regime (Fig. \ref{fig:QAZE} (a), (d)). In the regime
of large $\Gamma$ (short bath correlation time) and small $g$ (weak
coupling), as in Fig. \ref{fig:QAZE} (d), all the curves for different
$N$ tend to overlap with one another, and the QZE and QAZE behaviors
approach to those of the KKA.

\section{Conclusion}

In summary, we have investigated the influence of the bath memory
effect on the QZE and QAZE. The assumption of the bath state reset
to its original state after each instantaneous projective measurement
on the system in the traditional approach ignores equivalently the
cross-correlations of the bath operators at different Zeno intervals.
For measurement projected to a general initial system state and for
a bath with a considerable memory effect, the assumption is not valid.
To solve the dynamics, we derive an exact master equation for Lorentzian
bath which is suitable for the case that the qubit system undergoes
time-dependent non-unitary operations such as Zeno projections, and
we compare it with former methods for verification. Based on the exact
result we find that, in stark contrast to the behaviors found in the
KKA, the scaled average decay rates in unit Zeno interval $w_{n}$
in our GA display an oscillatory behavior enabling even in the regime
of very small Zeno intervals a non-exponential decay behavior in the
SP, and the total average decay rate depends not only on $\tau$ but
also on the number of repeated measurements $N$. For a fixed $N$,
some values of $\tau$ for which the traditional approach predicts
a QZE region may be in fact already in the QAZE region. Overall, the
width $\Gamma$ characterizes the damping rate of the memory and system-bath
coupling strength $g$ characterizes the memory depth of the bath.
So small $\Gamma$ and large $g$ make the cross-correlation between
different Zeno intervals substantially non-negligible, resulting in
both significant quantitative and qualitative differences between
the GA and KKA. Our results provide an essential step toward a further
in-depth and comprehensive understanding of the complex problems of
QZE and QAZE in open quantum systems. It will be interesting to see
whether our predictions can be verified experimentally in realistic
systems such as superconducting circuit QED systems.
\begin{acknowledgments}
Z.Z., Z.L. and H.Z. acknowledge support from the National Natural
Science Foundation of China under Grants No.~11374208 and No.~11474200.
H.S.G. acknowledges support from the the Ministry of Science and Technology
of Taiwan under Grants No.~103-2112-M-002-003-MY3 and
106-2112-M-002-013-MY3, from the National 
Taiwan University under Grants No. 105R891402 and No. 105R104021,
and from the thematic group program of the National Center for Theoretical
Sciences, Taiwan. 
\end{acknowledgments}


\begin{thebibliography}{10}
\bibitem{TA_SP_note} Another commonly seen formula through which
the SP $P_{\textrm{KKA}}\left(\tau\right)$ right after a single projective
measurement ($N=1$) is calculated in many QZE and QAZE studies in
the KKA is $P_{\textrm{KKA}}\left(\tau\right)={\rm Tr}_{S\otimes B}\left[\mathcal{P}_{S}U\left(\tau\right)\rho_{{\rm tot}}\left(0\right)U^{\dagger}\left(\tau\right)\mathcal{P}_{S}\right]$,
but then the assumption that the bath is reset to its initial state
$\rho_{B}\left(0\right)$ is implicitly made so that the total system-bath
evolution in the next Zeno interval is the same as that in the previous
interval and consequently $P_{\textrm{KKA}}\left(t\right)=\left[P_{\textrm{KKA}}\left(\tau\right)\right]^{N}$.
\bibitem{Derivation_Ps} The operator $e^{-\mu\left(I-\mathcal{P}_{S}\right)}=I+\sum_{n=1}^{\infty}\frac{\left(-\mu\right)^{n}\left(I-\mathcal{P}_{S}\right)^{n}}{n!}$.
Since $\left(I-\mathcal{P}_{S}\right)^{n}=I-\mathcal{P}_{S}$ for
$n\ge1$, one has $e^{-\mu\left(I-\mathcal{P}_{S}\right)}=\mathcal{P}_{S}+e^{-\mu}\left(I-\mathcal{P}_{S}\right)$.
So $\mathcal{P}_{S}=e^{-\mu\left(I-\mathcal{P}_{S}\right)}$ when
$\mu\to\infty$.
\end{thebibliography}
\end{document}